\documentclass[photonics,article,accept,moreauthors,pdftex,12pt,a4paper]{mdpi_mod} 
\lastpage{x}
\doinum{10.3390/------}
\pubvolume{}
\pubyear{}
\history{Submitted to MDPI Photonics, Bazel, Switzerland, on June 5, 2015}

\usepackage{graphicx}
\Title{Optical nanofibre integrated into an optical tweezers for particle manipulation, \textit{in-situ} fibre probing, and optical binding studies$^\dagger$}

\Author{Ivan Gusachenko $^{1}$, Viet Giang Truong$^{1}$, Mary C. Frawley$^{1, 2}$ and S\'ile Nic Chormaic$^{1,}$*}

\address{%
$^{1}$ Light-Matter Interactions Unit, Okinawa Institute of Science and Technology Graduate University, Onna, Okinawa 904-0495, Japan\\
$^{2}$ Physics Department, University College Cork, Cork, Ireland}
\corres{ E-mail: sile.nicchormaic@oist.jp; Tel.: +81-98-966-1643; Fax: +81-98-966-1063}

\abstract{Precise control of particle positioning is desirable in many optical propulsion and sorting applications. Here, we develop an integrated platform for particle manipulation consisting of a combined optical nanofibre and optical tweezers system. Individual silica microspheres were introduced to the nanofibre at arbitrary points using the optical tweezers, thereby producing pronounced dips in the fibre transmission. We show that such consistent and reversible transmission modulations depend on both particle and fibre diameter, and can be used as a reference point for \textit{in-situ} nanofibre or particle size measurement. Thence, we combine scanning electron microscope (SEM) size measurements with nanofibre transmission data to provide calibration for particle-based fibre assessment. This integrated optical platform provides a method for selective evanescent field manipulation of micron-sized particles and facilitates studies of optical binding and light-particle interaction dynamics.}


\conference{Proc. SPIE 9164, Optical Trapping and Optical Micromanipulation XI; \textit{Ivan Gusachenko, Mary C. Frawley, Viet. G. Truong and S\'ile Nic Chormaic},
"Optical nanofiber integrated into an optical tweezers for particle manipulation and in-situ fiber probing", 91642X (September 16, 2014); doi:10.1117/12.2061442}

\begin{document}


\section{Introduction}

A decade after its first realisation, Ashkin~\citep{Ashkin1970} used a laser to move micron-sized dielectric particles. Nowadays, light-mediated trapping boasts a myriad of experimental applications, including cell manipulation~\cite{Ashkin1987}, force measurement~\cite{Kuo1993}, holographic trapping~\cite{Curtis2002}, angular momentum transfer~\cite{Simpson1997}, and the use of optical fibres as trapping sources~\cite{Liu2006}. Recent advances include the study of light-matter coupling in optically bound structures~\cite{Dholakia2010}, and lab-on-a-chip integrated techniques~\cite{Enger2004}. A sister branch of optical manipulation that is subject to increasing attention exploits the potential of surface evanescent fields. Prism surfaces~\cite{Kawata1992}, microscope objectives~\cite{Gu2004}, and waveguides~\cite{Yang2009} have been used as evanescent field interfaces to probe the dynamics of optical trapping and binding, including extensive studies of counter propagating and standing-wave field effects~\cite{Siler2006, Siler2008}.

A special case of evanescent field geometry is found in the optical nanofibre (ONF). With a diameter comparable to the wavelength of light guided within, such ultrathin optical fibres have intense evanescent fields which penetrate into the surrounding medium~\cite{Tong2004}. Being relatively easy to fabricate and integrate with other optical components, nanofibres have emerged as compact, versatile devices with a broad range of applications~\cite{Tong_book}, such as cold-atom manipulation~\cite{Kumar2015} and microresonator coupling~\cite{Ward2008}.  For a review on some of this work, the reader is referred to ~\cite{Morrissey2013}.   Optical manipulation using nanofibres is now a field of increasing potential~\cite{Brambilla2007}; the evanescent field around the waist of the nanofibre is used to optically trap and propel micron-sized particles in suspension. In a manner reminiscent of conventional optical tweezers (OT), the gradient of the evanescent field attracts nearby particles to the fibre surface. These are then propelled along the direction of light propagation via radiation pressure. Since their establishment as optical propulsion tools~\cite{Brambilla2007}, nanofibres have been used for bidirectional particle conveyance~\cite{Lei2012}, wavelength selective particle sorting~\cite{Zhang2013}, and mass biological particle migration under photophoresis~\cite{Lei2011}. Such methods have exciting applications as particle 'conveyor belts' and sorting mechanisms in enclosed systems, particularly as their mm-scale lengths also facilitates continuous and long range trapping at any point in a sample, beyond limits achievable with conventional focussed-beam tweezers.

In most cases, nanofibres are immersed in a 'particle bath' -- a relatively high-density solution which allows many particles to simultaneously interact with the fibre. However, particles moving in and out of the evanescent field can cause major scattering-induced system fluctuations, making it difficult to determine the evanescent field incident on particles within a given visual frame. Microfluidic insertion is one way to address this problem~\cite{Xin2013}. Although such systems are highly relevant for mass particle sorting and filtration, they can be complex to arrange and have limitations in terms of particle/site selection and system reversibility.

Here, we introduce an integrated platform for particle manipulation using a combined optical nanofibre and optical tweezers system. This allows particles to be selectively trapped -- individually or in arrays -- and site-specifically introduced to the nanofibre surface. Previously we showed that addressing nanofibres in a more structured and site-specific way offers many advantages, including facilitating the study of interparticle or particle plus evanescent field interactions~\cite{Frawley2014, Maimaiti2015}.

Previously reported works on ONF-based particle manipulation rely solely on the video sequence analysis for studying particle displacement. However, as in a standard optical tweezers, the monitoring of the trapping beam affords new capabilities for the system such as position tracking~\cite{Gittes1998} or force measurement~\cite{Thalhammer2015}.  The ONF transmission can also be used as a relatively simple tool to provide complementary information about both trapping or trapped entities. The reason it has not been considered before probably stems from the difficulties associated with high colloidal particle concentrations in 'particle baths', as discussed earlier.

Taking advantage of our selective control over particle deposition onto ONFs, in this paper we implement ONF transmission monitoring to create an experimental toolbox for a range of trapping and manipulation studies. More precisely, we show that a single silica particle of known diameter can serve as a probe to sense the local fibre diameter by observing the nanofibre transmission. We also report ONF transmission as a function of particle-ONF separation, as well as preliminary results on the way the optical binding between particles on an ONF is related to its transmission. Our integrated technique may be further extended to provide a powerful, system-defined particle selection and manipulation tool, with broad spectroscopic applications analogous to those demonstrated in free space via injection of quantum dots onto the fibre surface~\cite{Yalla2012} or selective deposition of fluorescent microparticles in a Paul trap~\cite{Gregor2009}. The details and applications of this combined system are outlined below.


\section{Experimental Methods}


\subsection{Optical tweezers}

\begin{figure}[htbp]
\centering
\includegraphics[width=0.6\textwidth]{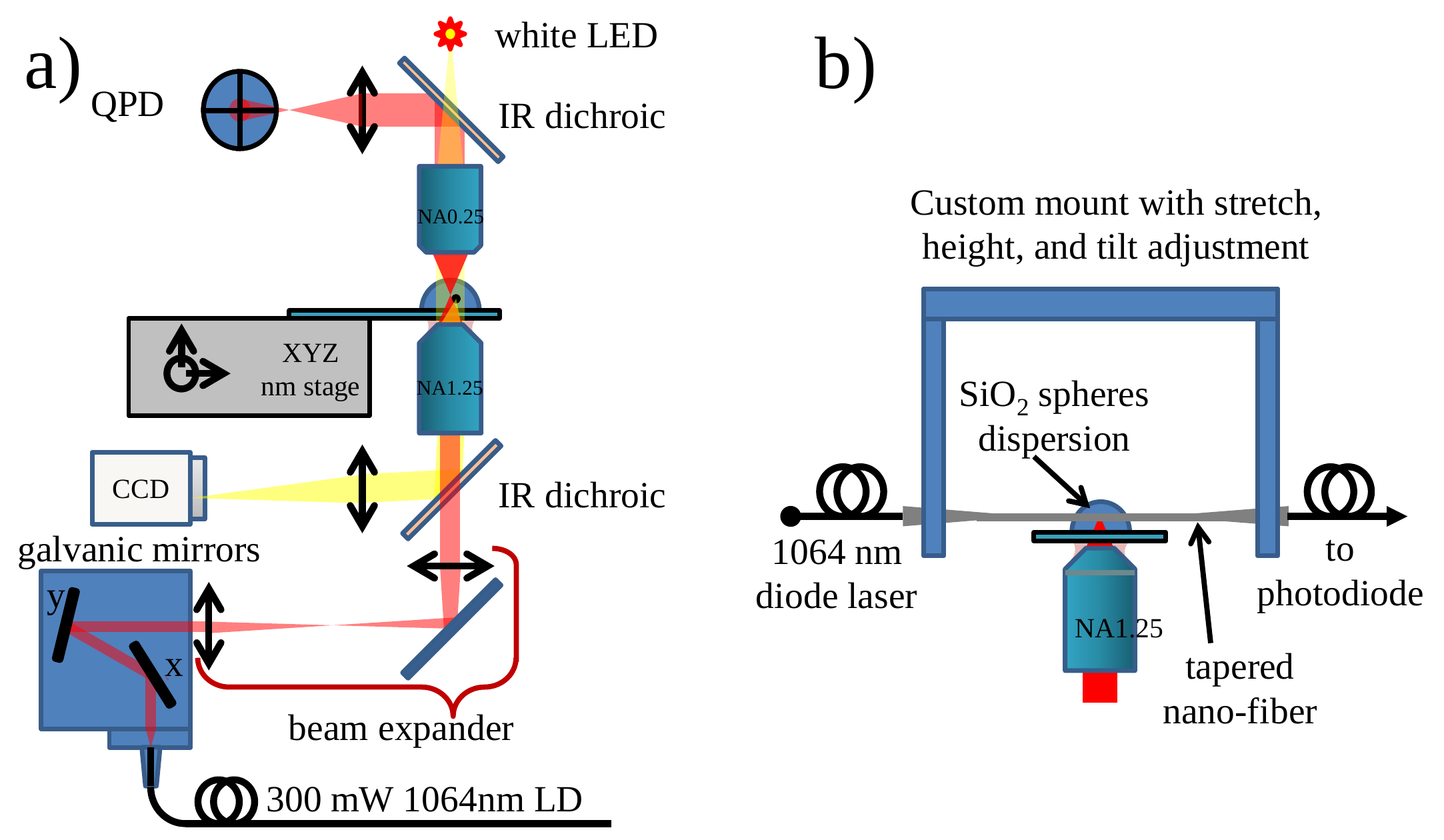}
\caption{\label{fig:opt_setup} a) Schematic of the optical tweezers setup with galvanic mirror steering of time-shared traps. b) The nanofibre is integrated into the optical tweezers using a custom-designed mount enabling the fibre to be positioned carefully in the microparticle solution.}
\end{figure}

To create the integrated tweezers-nanofibre system, we used a home-built tweezers (see Fig.~\ref{fig:opt_setup}(a)) based on Thorlabs model OTKB/M, with a 300~mW 1064~nm laser (hereafter referred to as the tweezers laser). It features an inverted microscope configuration, where the sample is placed over an oil-immersion objective (100x 1.25 NA) and illuminated from above. The tweezers includes a galvo-steered mirror pair in the incident beam path and this allows modulation of the beam position in the focal plane for simultaneous trapping of multiple particles through time-sharing of the beam.

\subsection{Optical nanofibre fabrication, mounting and integration}

The most common means of nanofibre fabrication involves stripping, heating and stretching standard optical fibre, until its waist region reaches nanoscale dimensions. Tried and tested heat sources include CO$_2$ lasers~\cite{Ward2006}, ceramic microheaters~\cite{Shi2006}, and butane or hydrogen torches~\cite{Birks1992}. For basic taper production, a modest pulling rig can easily be assembled in the laboratory -- the heat source (typically a few mm wide) remains stationary, while two linear stages draw the fibre outwards. This yields exponential taper shapes and high quality nanofibres with adiabatic transmissions close to 100\%~\cite{Love1991}. Introducing a lateral scanning function to the heat source immediately improves the rig functionality, potentially allowing the user to create and reproduce high quality nanofibres with arbitrary waist lengths, diameters, and taper profiles~\cite{Birks1992}.

\begin{figure}[htbp]
\centering
\includegraphics[width=0.6\textwidth]{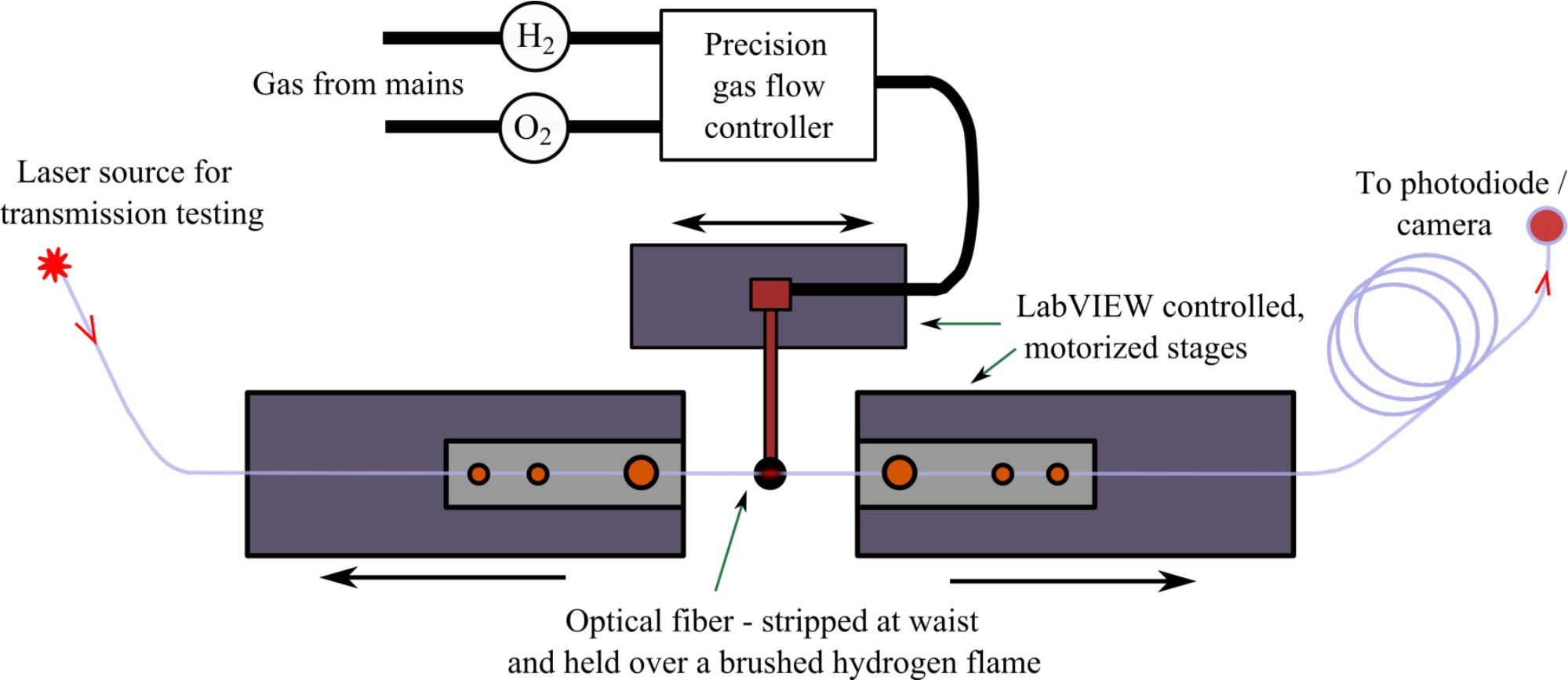}
\caption{\label{fig:fiber_rig} Schematic of a hydrogen:oxygen flame-brushed fibre pulling rig. The fibre is pulled apart by motorised stages, while being heated by a brushed hydrogen:oxygen flame. The transmission through the fibre is monitored during the pull.}
\end{figure}

Using a hydrogen:oxygen flame-brushed, heat-and-pull method~\cite{Ward2014} (see Fig.~\ref{fig:fiber_rig}(b)), optical nanofibres with waist diameters of approximately 530~nm were fabricated from Thorlabs 1060XP fibre. The nanofibres had linear tapers and waist lengths of 29 mm and 2 mm, respectively, and measured transmissions greater than 90\%. We selected a nanofibre diameter of 530~nm to satisfy the single-mode condition for 1064~nm~\cite{Tong2004} guided light and to ensure an enhanced evanescent field, while maintaining fibre robustness for handling. After that, the fibre was fixed to a custom mount, which was then firmly attached to the tweezers' 3D stage, positioning the nanofibre centrally over the pre-mounted coverslip (see Fig.~\ref{fig:opt_setup}(b)).


To facilitate stable co-planar focussing with the optical trap (1.25 NA objective with relatively small working distance), the nanofibre should be located within 100 $\mu$m of the sample coverslip surface due to the limited working distance of the objective. The tilt function of the mount allows the user to align the fibre parallel to the surface of the coverslip; this is important  both for even focussing of the fibre over the field of view, and to avoid contact between the fibre and the coverslip which results in complete loss of the fibre-guided light.

\subsection{Nanofibre in dispersion}

Silica microspheres of 2.01 $\mu$m and 3.13 $\mu$m (SS04N and SS05N Bangs Laboratories, Inc.) were diluted with ultrapure water (Nanopure, Thermo Scientific) to concentrations of $10^5-10^6$ $\mu$l$^{-1}$. Prior to positioning the fibre, 100 $\mu$l of the dispersion was pipetted onto the centre of the coverslip, and left for five minutes to allow the beads to settle to the bottom. Once mounted, one fibre pigtail was spliced to a 750~mW fibred 1064~nm diode laser (Oclaro), hereinafter referred to as the guided laser, and the transmitted power was monitored at the fibre output. The fibre was then lowered until it was just above the coverslip (see Fig.~\ref{fig:opt_setup}(b)). Finally, a second coverslip was positioned above the sample using short supports. This stabilises the system by alleviating atmospheric airflow across the droplet and serves to delay water evaporation, a cause of power fluctuations. With an exponentially-tapered fibre, the fibre transmission gradually increased upon lowering into the solution, until it reached a typical level of 80\% near the coverslip surface. The loss in this instance is repeatable, and largely due to scattering at the air-water interfaces of the thin fibre. However, if linearly-shaped tapers are used, the transmission is restored nearly to its initial value when the nanofibre is positioned close to the coverslip. This may be attributed to the larger fibre diameter as it enters and leaves the droplet in this geometry. Thus,  linear tapers prove to be more suitable for this type of application due to a more uniform waist and increased robustness.

\subsection{Data acquisition and treatment}

Once the nanofibre had been brought into the microparticle dispersion, single silica microparticles were trapped on the surface of the optical tweezers sample holder coverslip. The sample stage was then lowered to bring the trapped particles into common focus with the nanofibre. Next, spheres were introduced to the evanescent field of the fibre, by either manually translating the stage, or displacing the optical trap via galvanic mirror steering. The optical transmission was recorded from the analogue output of a Thorlabs PM100 power meter using a NI BNC-6229 DAQ card using Matlab-written software. The sampling rates were either 1 or 2 kHz.

\begin{figure}[htbp]
\centering
\includegraphics[width=0.6\textwidth]{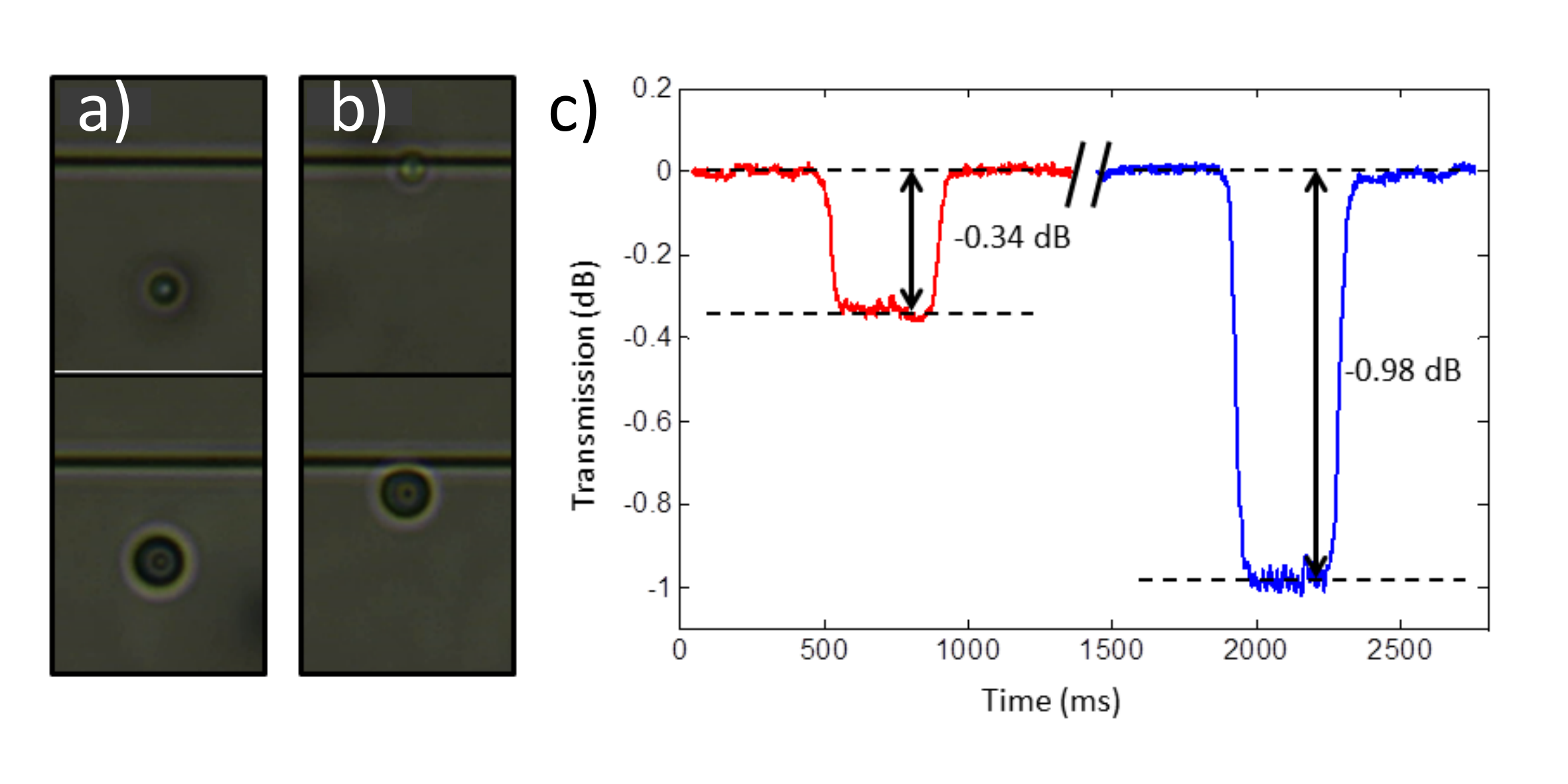}
\caption{\label{fig:2um_3um_approach} 2.01 $\mu$m (top) and 3.13 $\mu$m (bottom) silica microspheres trapped (a) near to and (b) in selective individual contact with the nanofibre surface. (c) Guided laser transmission dips due to particle scattering in the nanofibre evanescent field. Left (red) plot for 2.01 $\mu$m sphere, and right (blue) plot for 3.13 $\mu$m sphere show drastically different induced loss in transmission. }
\end{figure}

Figure~\ref{fig:2um_3um_approach} shows individually trapped particles in the vicinity of the nanofibre, with a 2.01 $\mu$m sphere in the top images and a 3.13 $\mu$m sphere in the bottom images. The trapped particles were introduced into the evanescent field of the fibre and sharp drops in transmitted power were consistently observed due to particle presence. The presence of a dip indicates that a portion of the evanescent field is coupled to the scatterer (sphere) and then radiated into the far field. Figure~\ref{fig:2um_3um_approach}(c) shows typical fibre transmission plots recorded for both particle sizes, with markedly larger transmission drops recorded for the larger sphere at the same fibre position. 

Before placing the fibre into the optical tweezers system, the 1064~nm guided light power was stable within 0.1\%. Once placed in the dispersion, the guided laser power fluctuated by 0.3 to 0.5\%. This increase in the fluctuations appears to arise from systemic noise in the combined tweezers-nanofibre system. To improve SNR and record nanofibre transmission changes over several repetitive events (5-15), the particle was approached to the fibre in a periodical manner using a Matlab program controlling galvanic mirrors. The recorded waveforms were then treated using  Matlab scripts to extract the best estimation for the transmission changes.

To exclude possible artifacts, it has been verified that even at OT trap powers >100~mW there was negligible coupling of light from the trap to the nanofibre (a few microWatts) whether a particle was present in the trap or not. Hence, the OT beam did not affect the transmission data.


\section{Results and Discussion}

\subsection{Trapped particle as a probe of fibre diameter and field distribution}

\begin{figure}[htbp]
\centering
\includegraphics[width=1\textwidth]{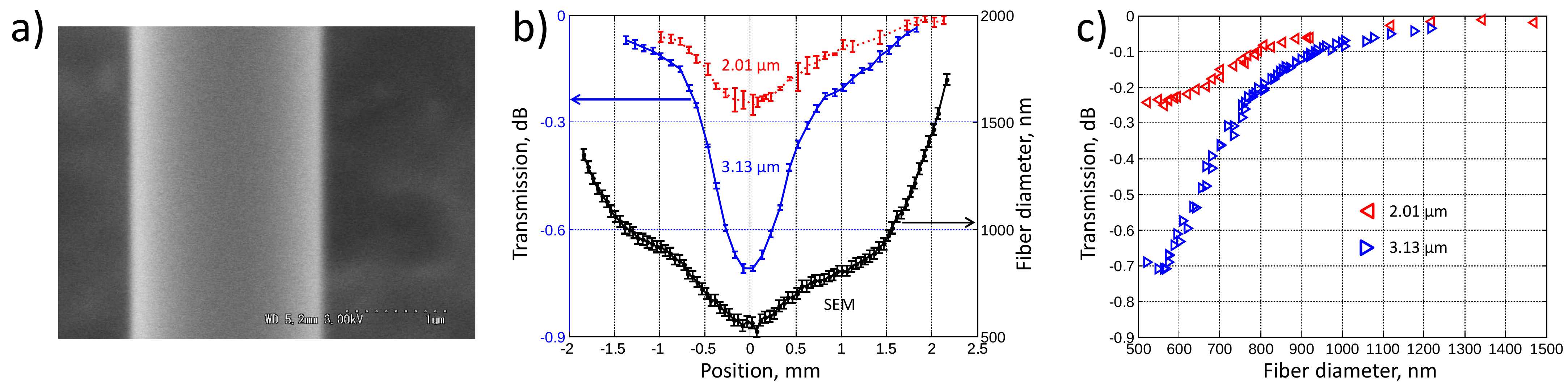}
\caption{\label{fig:sem_dip_calib} (a) Typical SEM image of the optical nanofibre. (b) Fibre transmission when in contact with a 2.01 $\mu$m (red line) and 3.13 $\mu$m (blue line) silica sphere, and the fibre diameter as measured from SEM images (black line), as a function of position along the fibre axis. (c) Fibre transmission as a function of fibre diameter when probed by silica particles with diameters of 2.01 $\mu$m (red left-pointing triangles) and 3.13 $\mu$m (blue right-pointing triangles).}
\end{figure}

As shown in Fig.~\ref{fig:2um_3um_approach}, the transmission dip induced by particles is dependent on particle size, probably due to the varying overlap of the fibre evanescent field with the particle. Since for a given wavelength a different fibre diameter results in a different evanescent field structure~\cite{Tong_book}, it is natural to consider the influence of nanofibre diameter on fibre transmission loss. Similar transmission effects to those in Fig.~\ref{fig:2um_3um_approach}(c) were systematically recorded for 2.01 $\mu$m and 3.13 $\mu$m particles, at 100 $\mu$m intervals along the fibre -- starting in the down taper region, continuing across the nanofibre waist, and finally along the up taper. The fibre profile was then measured via scanning electron microscopy (SEM) (Fig.~\ref{fig:sem_dip_calib}(a, b)) and the results correlated (see Fig.~\ref{fig:2um_3um_approach}(c)). The decreasing nanofibre diameter induces dramatic and increasing scattering in the particle; this scattering also increases with particle size. This effect can be associated with varying evanescent field penetration depths and intensity profiles at different fibre diameters for a given wavelength. These results provide a useful tool for \textit{in situ} measurements of nanofibre diameters, which is otherwise time consuming and can often only be measured afterwards using an SEM. Experimental knowledge of this relationship is thus important for emerging techniques for site-specific particle detection in low-density solutions, for instance, in fibre propulsion and optical binding studies~\cite{Brambilla2007, Frawley2014}.

\subsection{Fibre transmission as a function of fibre-particle separation}

While bringing particles to a fibre, in our case, resulted in sharp transmission modulation, it is expected that the modulation depth varies continuously with the fibre-particle separation. At the same time, empirical knowledge of this dependence can be helpful, first, to study the field structure of the fibre with a given particle, and second, to implement a high sensitivity probe for the particle-fibre separation. Thus, we also studied nanofibre transmission loss as a function of distance between the fibre and a probe particle of 3.13 $\mu$m diameter.

\begin{figure}[htbp]
\centering
\includegraphics[width=0.8\textwidth]{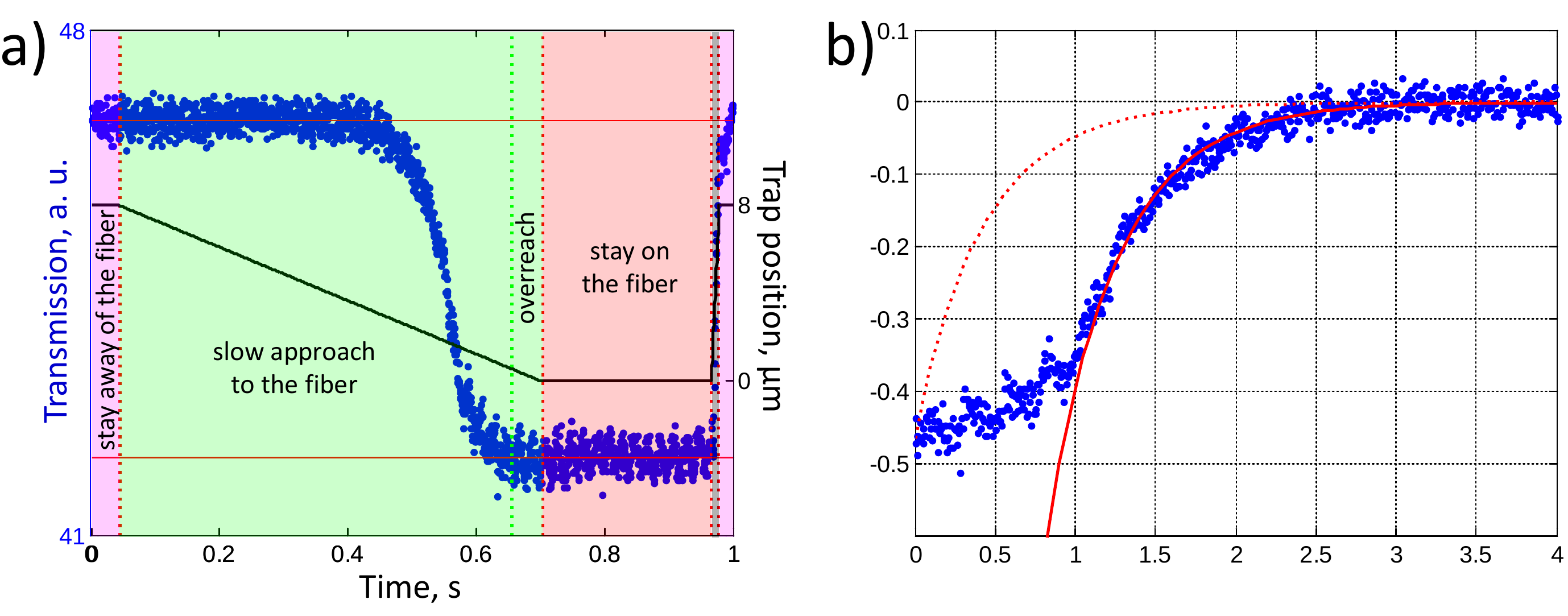}
\caption{\label{fig:fiber_part_distance} (a) Temporal profile of fibre transmission change (blue dots) during a cycle of optical trapping with a 3.13~$\mu$m sphere approaching and moving away from the fibre (black line). The trap movement is decomposed into four parts: (i) slow approach to the fibre (green zone), (ii) particle on the fibre (red zone), (iii) quick removal from the fibre (very thin gray zone), and (iv) particle at 8~$\mu$m distance away from the fibre (purple zone). Overreach zone corresponds to the particle being already at the surface of the fibre, while the trap continues to move in. (b) Fibre transmission as a function of fibre-particle distance, showing strongest dependence at 1-1.5~$\mu$m away from the fibre surface (blue dots); model fit matching the transmission at zero separation between fibre and particle (red dotted curve); model fit matching the experimental transmission at a distance from the fibre (solid red curve).}
\end{figure}

To perform the measurements, the optical trap displacement was performed in a periodic way, and fibre transmission was averaged over 10 consecutive periods, similar to what an averaging mode of a digital oscilloscope would produce. The resulting transmission profile is shown in Fig.~\ref{fig:fiber_part_distance}(a) as blue dots. The black line shows the trap position as a function of time within a 1 second-long period, which consisted of four distinct parts with corresponding relative time portion given in the brackets: slow approach to the fibre from 8~$\mu$m distance (5);  on the fibre (2); quick removal from the fibre (0.1); separated from the fibre by 8~$\mu$m  (0.5). Due to manual positioning of the oscillating trap with respect to the ONF, there was typically an overreach, i.e. the trap moved a little further after the sphere was brought in contact, making it slide azimuthally along the surface of the fibre. This overreach can be deduced from the imaging and corrected for.

The transmission versus separation curve was deduced from the part corresponding to the approach (Fig.~\ref{fig:fiber_part_distance}(b)). The dependence shows a smooth sigmoidal transition with highest sensitivity for fibre-particle distances between 0.5 and 2~$\mu$m. As can be seen from the figure, the sensitivity ($\partial T/\partial d$) is low for distances both smaller than the diameter of the fibre, and larger than the diameter of the particle.

To fit the experimental dependence, we exploit an intuitive yet simplified theoretical estimation for the transmission loss. In this model, we assume that the loss is proportional to the intensity of the induced dipole represented by the particle, which we estimate as the square of the field amplitude overlap with the sphere:

\begin{equation}
T = 1 - \gamma \left(\int E(r) dV \right)^2,
\end{equation}
where $\gamma$, being the free parameter of the model, is the proportionality factor between the transmission loss $\Delta T$ and the field overlap integral squared. The field radial profile, which is approximately exponential with a decay length of $\approx$0.88~$\mu$m, was obtained numerically from COMSOL simulations.

The fit that matches the measured transmission at the contact between fibre and particle cannot describe the highest sensitivity obtained at a significant distance from the fibre (Fig.~\ref{fig:fiber_part_distance}(b), dotted curve). The slow transmission variation at large distances (>1$\mu$m) is, however, consistent with this simple model, provided that we adjust the free parameter $\gamma$ to achieve the fit (Fig.~\ref{fig:fiber_part_distance}(b), solid curve). On the other hand, the model fails to explain the observed sigmoidal behaviour, and predicts excessively low transmission - around -7 dB - at the nanofibre surface. At the same time, this phenomenon can be explained considering that, for the light coupled from the fibre into the sphere, there is an interplay between scattering into the far field (transmission loss), and coupling back to the propagation mode of the fibre (no loss in transmission). In this case, our observations may imply that for an approaching particle, the increasing overlap with the evanescent field is compensated for by an increase of coupling back to the fibre and this alleviates further transmission decrease.

The obtained results suggest that ONFs can be used for colloidal particle sensing in  flow, with particles not necessarily touching the ONF surface. Besides providing the effective sensing volume of a single ONF for a given particle size, the separation-transmission dependence could be  useful for profiling  electromagnetic fields which have complex profiles, e.g. for higher order modes propagating in an ONF~\cite{Frawley2012, Maimaiti2015}. We also show that the observed, unexpected transmission dependence at small separations cannot be directly explained by the simplistic model we employed, which relies on field overlap integral, and this clearly requires further investigation.

\subsection{Transmission modulation induced by two particles}

\begin{figure}[htbp]
\centering
\includegraphics[width=0.5\textwidth]{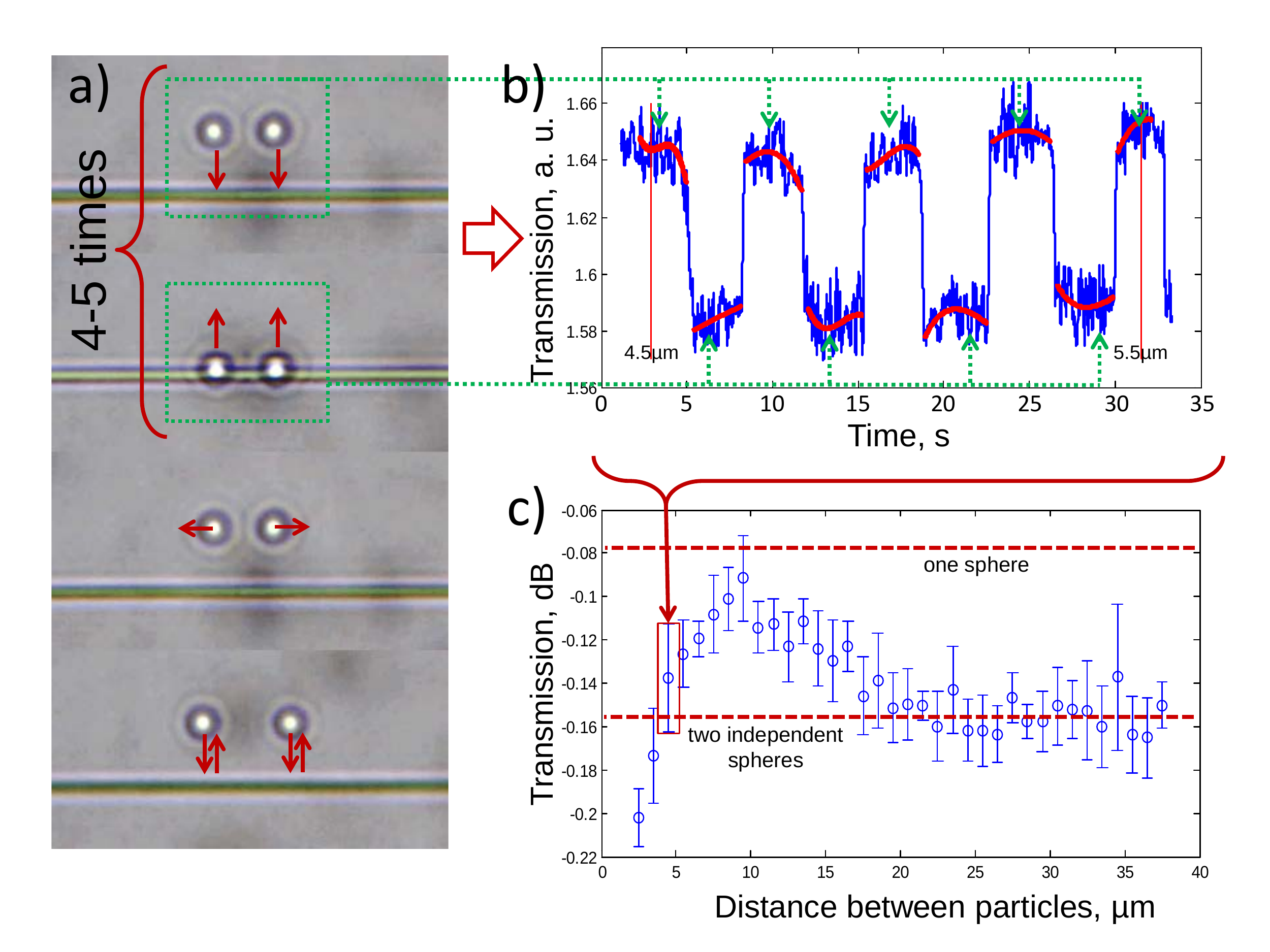}
\caption{\label{fig:2part_dip} Transmission modulation of a ~850~nm ONF induced by two 2.01~$\mu$m silica particles in contact. (a) Schematic of the experiment showing the succession of steps. (b) ONF transmission as a function of time shows reproducible dips when particles are brought into contact with the ONF. Red lines show the polynomial fit of the transmission. (c) Transmission loss for two particles on a fibre exhibits strong variation with the interparticle distance, indicating optical interaction. The red lines indicate the estimates for single-particle induced loss, as well as twice this value, corresponding to two independent, non-interacting particles.}
\end{figure}

In order to gain an insight into ONF-based optical binding~\cite{Frawley2014} via fibre transmission monitoring, we recorded the fibre transmission loss  on bringing two particles simultaneously into contact with a fibre. Two 2.01~$\mu$m diameter spheres were trapped in two time-shared optical traps, brought manually to the fibre (Fig.~\ref{fig:2part_dip}(a)), and the corresponding transmission changes were recorded and analysed (Fig.~\ref{fig:2part_dip}(b)).

We observed strong transmission dependence on the interparticle distance (Fig.~\ref{fig:2part_dip}(c)). It is meaningful to compare the recorded transmission loss with twice the loss induced by a single particle (red lines in Fig.~\ref{fig:2part_dip}(c)),  corresponding to the absence of any optical interaction between particles on the fibre. The recorded experimental curve consists of values which can be both smaller and larger than the indicated benchmark value of two independent particles; this clearly indicates the presence of an optical interaction.

The optical binding on an ONF can be considered in two different geometries, using either a single propagation direction~\cite{Frawley2014}, or counter propagating beams~\cite{Dholakia2010}. In  particle sorting applications that are particularly suited for ONFs, the binding within propagating clusters is what is relevant. Experimentally, this involves tracking fast moving particles over distances larger than several typical fields of view (FOV) of a microscope, which is often impractical. Provided that the link between ONF transmission modulation and particle binding is established, current results open possibilities for studying such  optical binding for propagating clusters using a stationary trap system, within a single FOV. However, additional experiments explicitly evaluating binding together with ONF transmission are required.


\section{Conclusions}

We have presented a combined optical nanofibre-optical tweezers integrated platform for particle propulsion and manipulation. Notably, we studied fibre transmission variations as a function of sphere size, fibre diameter, fibre-sphere separation, and demonstrated the particular effect of transmission modulation for two particles on a fibre. Our system presents a powerful framework for studying various interactions between particles and the evanescent field of the fibre, while also serving as a tool for other fibre-based particle propulsion and sorting experiments.  In particular, it enables us to determine  local fibre diameter \textit{in situ}, as well as potentially elucidating optical binding behaviour in such systems.

The tweezers and nanofibre intensities can be modulated to trap at a specific surface position on the fibre, or propel in both directions by balancing bidirectional nanofibre fields. This setup is also perfect for trapping multiple particles, and was successfully used for studying optical binding interactions occurring in the evanescent field of an optical nanofibre~\cite{Frawley2014}.

Although the experiments described above used silica microspheres, the techniques are also applicable to optical interaction studies and manipulation of other dielectric or biological specimens, notably polystyrene spheres and living cells. Extended functionality of the tweezers will introduce further experimental possibilities.  For example, introducing high precision calibrated particle tracking via a quadruple photo diode~\cite{Gittes1998} or a high-speed camera~\cite{Gibson2008} would enable site-specific force measurements of the evanescent field mediated particle-particle and fibre-particle interactions.

Finally, the nanofibre field may be tailored by changing the input wavelength, power, or light polarisation; this shows that the combined tweezers-nanofibre system  has further potential for use in fluorescence and spectroscopy studies, using the nanofibre as a passive (collection) or active (excitation) probe. The evanescent field structures of non-uniform fibre elements~\cite{Xin2011}, and complex evanescent fields under counter propagating light~\cite{Lei2012} and higher mode propagation~\cite{Frawley2012, Ravets2013, Maimaiti2015} may also be investigated using this system.


\acknowledgments{Acknowledgments}

This work was supported in part by funding from the Okinawa Institute of Science and Technology Graduate University and and Science Foundation Ireland under
Grant No. 08/ERA/I1761 through the NanoSci- E + Transnational Programme.  


\authorcontributions{Author Contributions}

IG planned and conducted the experiment, treated data, and wrote and prepared the manuscript. MF participated in writing the manuscript. VGT participated in planning the experiment, performed COMSOL simulation, and edited the manuscript. SNC proposed and supervised the project, and edited the manuscript.


\conflictofinterests{Conflicts of Interest}

The authors declare no conflict of interest. 

\bibliographystyle{mdpi}
\makeatletter
\renewcommand\@biblabel[1]{#1. }
\makeatother






\bibliography{ONF_TW_BIB}
\bibliographystyle{mdpi}


%


%

\end{document}